\documentclass{iopart}
\usepackage{graphics,multicol}
\usepackage[a4paper,margin=3cm,lmargin=3cm,headheight=1pt]{geometry}
\usepackage{fancyhdr}
\usepackage{amssymb}

\def\cF{\mathcal{F}}
\def\bs{{\bf s}}
\def\va{V(a)}
\def\ind{{\mathbb I}}

\begin{document}

\paper{The number of matchings in random graphs}

\author{Lenka Zdeborov\'a and Marc M\'ezard}
\address{CNRS; Univ. Paris-Sud, UMR 8626, LPTMS, ORSAY CEDEX, F-91405
}
\ead{\mailto{zdeborov@lptms.u-psud.fr}, \mailto{mezard@lptms.u-psud.fr}}

\begin{abstract}
We study matchings on sparse random graphs by means of the cavity method. 
We first show how the method reproduces several known results about maximum 
and perfect matchings in regular and Erd\"os-R\'enyi random graphs. 
Our main new result is the computation of the entropy, i.e. the 
leading order of the logarithm of the number of solutions, 
of matchings with a given size. 
We derive both an algorithm to compute this entropy for an arbitrary graph 
with a girth that diverges in the large size limit, and
an analytic result for the entropy in regular and Erd\"os-R\'enyi 
random graph ensembles.
 
\end{abstract}

\section{Introduction}

We study a classical problem of graph theory, namely the size and number of
matchings on various types of random graphs. This problem has been intensively
studied for a long time by mathematicians and computer scientists
\cite{Lovasz}. Here we address it using some techniques developed in the
statistical mechanics of spin glasses \cite{MPbook}. Such approaches have been
used in recent years to describe successfully the typical cases of random
combinatorial problems as e.g. the weighted matching (or assignment)
\cite{MP85}, the traveling salesman problem \cite{MP86}, the vertex cover on
random graphs \cite{WH00}, K-satisfiability \cite{MPZ02, MZ02}, or the coloring
of random graphs \cite{MPWZ02}.

Here we apply the cavity method \cite{MP99} to describe the matchings on
ensembles of sparse random graphs with a given degree distribution. We work
within the replica symmetric (RS) version of the cavity method, and we argue
that it gives exact results for these problems. In fact we show how the method
reproduces several known results about the size of the maximum matching (which is also
the maximum number of self avoiding dimers)
and the existence of the perfect matchings (the possibility of covering the graph with $N/2$ dimers). 
This also confirms the previous result by Zhou and
Ou-Yang \cite{ZOY03} who also used the cavity method, but in a different way 
(we discuss below the differences of our approaches).

Our main
new result is the computation of the entropy, i.e. the leading order of the
logarithm of the number of solutions, of matchings with a given size in large
sparse random graphs. We derive both an algorithm to compute this 
entropy for arbitrary graphs with a girth (the length of the 
shortest graph cycle) that diverges in the large size limit, and
an analytic result for the entropy in regular and Erd\"os-R\'enyi 
random graph ensembles.

The cavity method is not yet proved to be a rigorous tool, however it is
widely believed -at least by physicists- to be exact, and in some cases its
predictions have been confirmed rigorously. Let us mention the work of
Talagrand \cite{Talagrand03} who, using some of the tools developed by Guerra
\cite{Guerra03}, proved the validity of the Parisi formula for the partition
function of Sherrington-Kirkpatrick model (Parisi's original
work \cite{Parisi80} uses the replica method, but it can be reformulated in
cavity terms \cite{MPbook}). Aldous \cite{Aldous01} developed the local weak
convergence method and proved the $\zeta(2)$-limit for the random assignment
problem, initially found in \cite{MP85}. In this same problem, Bayati, Shah
and Sharma \cite{BSS05} proved the convergence of a ``belief propagation''
algorithm, which is basically the replica symmetric cavity method, for finding
the lowest weight assignment in generic bipartite graphs. Recently, 
Bandyopadhyay and
Gamarnik \cite{BG05} have used this local weak convergence strategy to derive
some results on the entropies in the problems of graph coloring and
independent sets, in regions of parameters where the RS cavity solution is the
correct one. The local weak convergence method was used for weighted matchings
in sparse random graphs also in \cite{GNS05}.

Because of  these recent developments, and of the simple 
replica symmetric nature of the matching problem, we believe that it 
should be possible to turn all our results into rigorous statements.
We hope that this work will turn to be useful also in the opposite direction,
i.e. that  working on rigorous proofs of our results for matching
will help to develop the rigorous version of the cavity method.

The matching problem on a graph  is equivalent to a physical model of dimers. 
 This was mostly studied on planar graphs (lattices), where 
there is a beautiful method by Kasteleyn \cite{Kasteleyn} which 
shows how to count exactly dimer arrangements (perfect matchings).
On non planar regular graphs a Bethe mean field 
approximation, which is known to be exact on Bethe lattice,
has been developed in \cite{HC04}, and references therein. Our work 
generalizes these results and gives the solution of dimer models 
on sparse random graphs. 

The paper is organized as follows. In section \ref{not} we set up
our notations and overview the main known results for the matching 
on sparse random graphs. In section \ref{cav} we introduce the cavity 
approach to the matching problem and derive the size of maximum matching
and the entropy of matchings of a given size on a typical random graph. 
We also describe approximate polynomial algorithms for sampling and 
counting matchings on a given graph. In section \ref{results} we give 
results for the size and the number of matchings for the ensemble 
of regular and Erd\"os-R\'enyi random graphs, and we show that 
the replica symmetric ansatz is stable for these two ensembles. 
In section \ref{1RSB} we discuss the alternative 1RSB solution at zero 
temperature which  was obtained  by Zhou and 
Ou-Yang \cite{ZOY03}. The conclusion summarizes this work and gives
some perspective on how it could be turned into rigorous results.

\section{Background and notations}
\label{not}

Consider a graph $G(V,E)$ with $N$ vertices ($N=|V|$) and a set of 
edges $E$. 
A {\it matching} of $G$ is a subset of edges $M \subset E$ such  
that each vertex is incident with at most one edge in $M$.
In other words the edges in the matching $M$ do not touch 
each other. 
The {\it size of the matching}, $|M|$, is the number of edges in 
$M$. We denote the size of maximum possible matching by $|M^*|$.
The trivial relation $|M^*| \le N/2$ follows from the definition. 
If a maximum matching covers all the vertices, $|M^*| = N/2$, we call $M^*$ 
a {\it perfect matching}. 

Finding a maximum matching in a given 
graph $G$ is a polynomial problem. For instance the algorithm of 
\cite{MV80} solves this problem with a computational complexity 
proportional to $O(|E| \sqrt{|V|}) $. 

How many matchings of size $|M^*|$ can 
we actually find in $G$? No exact polynomial algorithm to answer this 
question is known.
Counting the number of matchings of a given size was proven 
\cite{Valiant79} to belong to the $\#P-$complete (sharp P-complete) 
class of problems. It means that if an exact polynomial algorithm
for this problem existed we could also count solutions of all 
the other problems belonging to the NP class. 
It is generally believed that a polynomial procedure to solve 
$\#P-$complete problems does not exist. 
For this reason it is very useful to develop methods to count matchings
fast (in polynomial time) but only approximately. Several works
have been done in this direction \cite{JS97, FK04, Chien04}. 

In this paper we study not only properties of matchings on a given graph 
$G$ but also on ensembles ${\cal G}$ of large sparse random graphs. 
When we claim that a property $A$ is true for a typical random 
graph $G\in {\cal G}$ we mean: when $G$ is chosen from the ensemble with its 
natural probability law, the probability that $A$ is true goes 
to one as the size of $G$ grows to infinity. 

\subsection{Rigorous results for matching on random graphs}

In this section we give some known rigorous results for matchings 
on random graphs. From the point of view of matching the simplest 
ensemble is the one of $r$-regular random graphs, i.e. all graphs where 
every vertex has degree (number  of neighbors)~$r$. 
In this ensemble the measure is uniform over all $r$-regular graphs. 

{\bf Theorem 1 (Bollob\'as and McKay'86 \cite{BM86}):} If $r\ge 3$ and
the number of vertices $N$ is even then almost every $r$-regular graph has a 
perfect matching. Denote by ${\cal N}_G$ the number of perfect matchings of 
$r$-regular graph $G$. Then its first two moments are 
\begin{eqnarray}
         \mathbb{E}({\cal N}_G) &\approx & \sqrt{2} e^{1/4} [(r-1)^{r-1}/r^{r-2} ]^{N/2}, 
                \label{reg_ann1}\\
         \mathbb{E}[({\cal N}_G)^2]  &\approx & \sqrt{\frac{r-1}{r-2}} 
      e^{-(2r-1)/4(r-1)^2}\, \mathbb{E}({\cal N}_G)^2 . \label{reg_ann2}
\end{eqnarray}

In the statistical physics language we call the logarithm of the 
first moment $\log{[\mathbb{E}({\cal N}_G)]}$ the {\it annealed entropy} 
of perfect matchings and the typical average $\mathbb{E}[\log{{\cal N}_G}]$ 
\footnote{We should write $\mathbb{E} [ \log{ ({\cal N}_G+1)} ]$ for 
the quenched entropy to avoid $-\infty$ for graphs which do not have 
any perfect matching.} 
the {\it quenched entropy} of perfect matchings. Due to the concavity 
of logarithmic function the upper bound for the quenched entropy 
follows from (\ref{reg_ann1}) 
\begin{equation}
   \mathbb{E}[\log{{\cal N}_G}] \le \log{[\mathbb{E}({\cal N}_G)]}= 
      N[(r-1)\log{(r-1)}-(r-2)\log{r}]/2 + O(1)\, . 
    \label{ent_reg}
\end{equation}
In section \ref{res_reg} we will show that for $r$-regular graphs 
the quenched entropy is in fact the same as the annealed one,
i.e. in (\ref{ent_reg}) the bound is tight. Note that the
fact that $\mathbb{E}[({\cal N}_G)^2]\sim [\mathbb{E}({\cal N}_G)]^2$
to leading exponential 
order is not enough to prove that the quenched and annealed entropy are equal. 

In this paper we will be interested in 
random graphs with a fixed degree distribution: we call ${\cal Q}(k)$
the probability that a randomly chosen vertex has degree $k$,
in the asymptotic limit of large graphs. In particular, in 
Erd\"os-R\'enyi 
(ER) random graphs,  where every edge is present with 
probability $p=c/(N-1)$, the degree sequence is Poissonian 
${\cal Q}(k)= e^{-c}c^k/k!$. Because of the existence of a 
fraction $e^{-c}$ of isolated  vertices perfect matchings
almost surely do not exist in ER graphs. 
The size of maximum possible matching was computed in a seminal paper of 
Karp and Sipser \cite{KS81}.

{\bf Theorem 2 (Karp and Sipser'81):} The maximum matching in 
an Erd\"os-R\'enyi random graph with $N$ sites and mean 
degree $c$ has on average size
\begin{equation}
     \mathbb{E}( |M^*| ) =  \frac{1-p_1(c)+p_2(c)-cp_1(c)+cp_1(c)p_2(c)}{2} N\, , 
     \label{exact}
\end{equation}
where $p_1(c)$ is the smallest solution of equation 
$p = \exp{[-c\exp{(-cp)}]}$ and $p_2(c)=1-\exp{[-cp_1(c)]}$. 

When $c < e$ there is only one solution for $p_1(c)$. When $c \ge e $ 
another pair of solutions for $p_1(c)$ appears.

\subsection{Karp-Sipser leaf removal, the core}

The Karp-Sipser theorem was originally proven by analyzing a 
greedy leaf removal algorithm \cite{KS81}. This algorithm consists 
of two steps
\begin{itemize}
  \item[(1)]{ Given a graph $G$, if there are leaves choose 
   randomly one of them $i$ and its incident edge 
   $(ij)$. Put this edge to the matching and remove 
   the two vertices $i$ and $j$. Delete at the same time 
   all the edges incident with $j$. Repeat until there are no leaves.}
  \item[(2)]{If there are no leaves in $G$ choose randomly an edge $(ij)$
   with uniform probability, add it 
   to the matching and erase all the edges incident with $i$ and $j$. 
   Go to step (1).}
\end{itemize}

We define as a {\it core of the graph} all the non-single vertices 
(and edges between them) which remain in the graph after the 
first step of the leaf removal procedure. The core does not depend on 
the history of the leaf removal \cite{BG01}. Karp and Sipser proved 
that for $c\le e$ the core is small (zero asymptotically) whereas 
for $c>e$ the core covers a finite 
fraction of all the vertices. They proved also that when a large
(or order $N$) core exists, asymptotically all its nodes can 
be matched. 

We call $v(c)$ the fraction of vertices in the core of 
a typical ER random graph of average degree $c$, 
$l(c)N$ the number of edges in the core, 
and $m(c)N$ the number of edges matched in 
the leaf removal procedure. It is known \cite{BG01} that 
\begin{eqnarray}
    v(c)= p_3 (1-cp_1)\, , \quad 
    l(c)= \frac{c}{2} p_3^2\, , \quad 
    m(c)= p_2 - \frac{c}{2} p_1^2\, , 
\end{eqnarray} 
where $p_1$, $p_2$ are the same parameters as in 
the Theorem 2 of Karp and Sipser, and $p_3=1-p_1-p_2$.
 
Properties of the core were also studied in \cite{APF98} and \cite{HW05}.
From these results it follows that the degree distribution in the core 
is Poissonian-like
\begin{equation}
  {\cal Q}(0)={\cal Q}(1)=0\, , \qquad 
  {\cal Q}(k)=\frac{e^{-cp_3} (cp_3)^k}{{\cal C} k!} \quad {\rm for} \quad k>1\, ,
   \label{core_deg}
\end{equation} 
where ${\cal C}$ is a normalization constant.
We will study connections between the Karp-Sipser 
leaf removal and our method in section \ref{core}.

\subsection{The annealed average}

We denote by ${\cal N}_G(x)$ the number of matchings of size $|M|=xN/2$ in 
a graph $G$. Its expectation $\mathbb{E}[{\cal N}_G(x)]$ in the random  
$r$-regular graph ensemble can be computed as the number of all possible 
matchings of size $|M|$ times the leading order 
of the number of all $r$-regular graphs
which contain a given matching $M$, divided by the leading order of the 
number of all $r$-regular graphs. We keep in mind that $r$ is finite and 
$N \to \infty$. This gives the annealed entropy: 
\begin{equation}
 \frac{\log{\mathbb{E}[{\cal N}_G(x)}]}{N} = \left(x-\frac{r}{2}\right)\log{r} 
 +\frac{r-x}{2}\log{(r-x)} -(1-x)\log{(1-x)}-\frac{x}{2}\log{x}\, .
  \label{ent_reg_ann}     
\end{equation}
Again, thanks to the concavity of logarithm, the quenched entropy 
cannot be larger than the annealed one $\mathbb{E}[\log{({\cal N})}]/N \le 
\log{\mathbb{E}({\cal N})}/N$.
We will see in section \ref{res_reg} that for $r$-regular 
graphs  this upper bound is actually tight. 

In the ensemble of ER random graphs the expectation of the number of 
matchings of size $|M|=xN/2$ is computed in the very same way and reads
\begin{eqnarray}
 \mathbb{E}[{\cal N}_G(x)] &\approx& 
             \exp{\left\{ \frac{Nx}{2} \left[ \ln{\frac{c}{x}} -1 -
             2\left(\frac{1}{x}-1\right)\ln{(1-x)}  \right]   \right\} }.
   \label{annealed}
\end{eqnarray}
If the exponent is negative (which happens for $c<e$ and $x$ 
sufficiently large) then there is almost surely no 
matching of size $|M|$ in graph $G$. On the other hand, if the exponent 
is positive then eq.~(\ref{annealed}) provides us with an upper bound 
on the quenched (typical) entropy
\begin{equation}
   \frac{\mathbb{E}\{\log{[{\cal N}_G(x)]}\}}{N} \le \frac{
   \log{\mathbb{E}({\cal N})}}{N} = \frac{x}{2} 
   \left[ \ln{\frac{c}{x}} -1 -
   2\left(\frac{1}{x}-1\right)\ln{(1-x)}  \right].
       \label{s_ann}
\end{equation}

For ER random graphs the bound is not tight. From eq.~(\ref{annealed}) 
we see that for $c > e$ the average number of perfect 
matchings ($x=1$) is exponentially large. But we know that for a typical 
ER graph no perfect matching exists (due to the presence of isolated 
vertices). The reason is that $\mathbb{E}[{\cal N}_G(x)]$ is dominated 
by few exceptional graphs $G$ which have a huge amount 
of perfect matchings. The correct quenched average 
$\mathbb{E}\{\log{[{\cal{N}}_G(x)]}\}/N$ will
be computed in section \ref{res_ER}.

\section{Cavity method: general formalism}
\label{cav}

\subsection{Statistical physics description}
We describe a matching by 
the variables $s_{i}=s_{(ab)} \in\{0,1\}$ assigned to each edge $i=(ab)$ of $G$, 
with $s_{i}=1$ if $i \in M$ and $s_{i}=0$ otherwise. 
The hard constraints that two edges 
in a matching cannot touch impose that, on each vertex $a\in V$: 
$\sum_{b, (ab)\in E} s_{(ab)} \le 1$. To complete our statistical
physics description, we define for each given graph $G$ an energy 
(or cost) function which gives, for each
matching $M=\{s\}$, the number of unmatched vertices:
\begin{equation}
   E_G(M=\{s\}) = \sum_{a} E_a(\{s\}) = N - 2|M|\ , 
   \label{Ham}
\end{equation}
where $E_a=1-\sum_{b} s_{(ab)}$. 
The Boltzmann probability law in the space of matchings
is defined by:
\begin{equation}
   P_G(M) = \frac{1}{Z_G(\beta)} e^{-\beta E_G(M)}\ ,
   \label{Boltzmann}
\end{equation}
where $\beta$ is the inverse temperature and $Z_G(\beta)$ is the
partition function.

\begin{figure}[!ht]
 \begin{minipage}{0.48\linewidth}
  \begin{center}
   \resizebox{5cm}{!}{\includegraphics{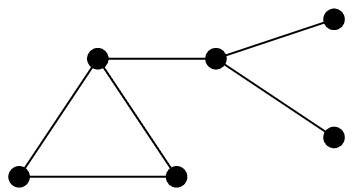}}
  \end{center}
 \end{minipage}
 \begin{minipage}{0.48\linewidth}
   \begin{center}
   \resizebox{5cm}{!}{\includegraphics{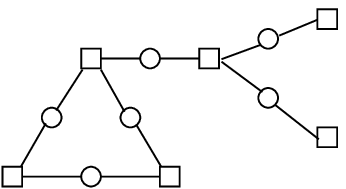}}
   \end{center}
 \end{minipage}
  \caption{\label{gr_fact} On the left, example of a graph with six nodes 
   and six edges. On the right, the corresponding factor graph with six functional 
   nodes (squares) and six variable nodes (circles). }
\end{figure}

We use a factor graph representation \cite{KFL01}
of the Boltzmann probability (\ref{Boltzmann}).  To a  graph $G$ we 
associate a factor graph $\cF(G)$ as follows (see fig. \ref{gr_fact}):
 To each edge of $G$ corresponds a
variable nodes (circle) in $\cF(G)$; to each vertex of $G$ corresponds a
function node (square) in $\cF(G)$. We shall index the variable nodes by
indices $i,j,k,\dots$ and function nodes by $a,b,c,\dots$. The variable $i$
takes value $s_i=1$ if the corresponding edge is in the matching, and $s_i=0$
if it is not. For a given  configuration
$\bs=\{s_1,\dots,s_{|E|}\}$, the weight of function node $a$ is
\begin{equation}
\psi_a(\bs)=\ind \left(\sum_{i\in\va}s_i\le 1\right) e^{-\beta(1-\sum_{i\in\va}s_i)}\ ,
\end{equation}
where $\va$ is the set of all the variable nodes which are neighbours of
function node $a$, and the total Boltzmann weight of a configuration
is $\frac{1}{Z_G(\beta)}\prod_a \psi_a(\bs)$. Later on, when 
confusion cannot be made, we denote $\va$ just as $a$. 

We want to compute the internal energy $E_G(\beta)$ (the expectation 
value of the number of unmatched vertices) and the entropy 
$S_G(\beta)$ (the logarithm of the number of matchings). 
For $\beta \to \infty$ (zero temperature limit) these two quantities give 
the ground state properties, i.e. respectively the size and entropy of 
the maximum size matchings. 

We are interested in the ``thermodynamic'' limit of large graphs ($N\to
\infty$), and we shall compute expectations over ensembles of graphs of the
densities of thermodynamical potentials $\epsilon(\beta)=
{\mathbb{E}}[ E_G(\beta) ]/N $ and $s(\beta)=
{\mathbb{E}}[S_G(\beta)]/N$, as well as the average free energy density
\begin{equation}
f(\beta)=\frac{-1}{\beta N}{\mathbb{E}}[\log{Z_G(\beta)} ]= \frac{1}{N}
{\mathbb{E}} [F_G(\beta)]  = \epsilon(\beta)- \frac{1}{\beta} s(\beta)\, .
\end{equation}
The reason for this interest is that one expects, for reasonable 
graph ensembles, $F_G(\beta)$ to be self-averaging. This means that 
the distribution of $F_G(\beta)/N$ becomes more and more sharply peaked 
around $f(\beta)$ when $N$ increases.

\subsection{The cavity method at finite temperature}

\begin{figure}[!ht]
  \begin{center}
   \resizebox{3cm}{!}{\includegraphics{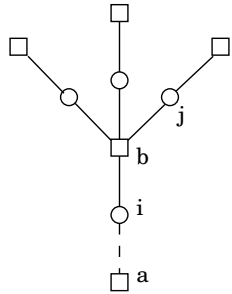}}
  \end{center}
   \caption{\label{fac_gr} Part of the factor graph used to compute 
   $P_{s_i}^{i\to a}$.}
\end{figure}
  

In the following we use the cavity method at the replica symmetric (RS) level,
as it is described in \cite{MP99}. We introduce a ``cavity'' in the factor 
graph by deleting the function node $a$ and its incident edges,
and we denote by $P_{s_i}^{i\to a}$ the probability that variable 
$i$ takes value $s_i$ (see fig. \ref{fac_gr}).  Because of the local 
tree-like structure of the (sparse) graphs
that we study, it is reasonable to assume that the $P_{s_j}^{j\to b}$ for $j
\in b - i$ are uncorrelated. This is the main assumption of the cavity method
at the RS level (see \cite{MP99}) and we will check later on its
self-consistency. Using this assumption one gets
\begin{eqnarray}
    P_{s_i}^{i\to a}&=& \frac{1}{{\cal C}^{i\to a}} \sum_{\{s_j\}}
    \ind\left(s_i+\sum_{j\in b - i} s_j  \le 1 \right)
    e^{- \beta(1-s_i-\sum s_j)}\, \prod_{j\in b - i}      
    P_{s_j}^{j \to b},  \label{cav_Z}
\end{eqnarray} 
where ${\cal C}^{i\to a}$ is a normalization constant.

For every edge between a variable $i$ and a function node $a$,  
we define a cavity field $ h^{i\to a}$ as 
\begin{equation}
    e^{-\beta h^{i\to a}} \equiv \frac{P_{0}^{i\to a} }{
    P_{1}^{i\to a} }\ .
    \label{def_cav_h}
\end{equation}
The recursion  relation between cavity fields is then:
\begin{eqnarray}
    h^{i\to a}= - \frac{1}{\beta} \log{\left[e^{-\beta}+ 
        \sum_{j\in b - i} e^{\beta h^{j\to b}} 
    \right]} \label{BP_T}.
\end{eqnarray}
This is one form of the ``belief propagation'' equations 
\cite{KFL01, pearl}.
The cavity fields  can be interpreted as messages living on the edges
of the factor graph, with some consistency rules on the function nodes,
and one can try to solve them by an iterative 
``message passing'' procedure. 

Assuming that one has found the cavity fields, one can deduce from them
the various marginal probabilities and the free energy.
For instance the expectation value (with respect to Boltzmann's 
distribution) of
the occupation number $s_i$ of a given edge $i=(ab)$ is equal to 
\begin{equation}
\langle s_i\rangle = \frac{1}{1+e^{-\beta(h^{i\to a}+h^{i\to b})}}\, .
\label{marginal}
\end{equation}
 
To compute the free energy we first define the free energy 
shift $\Delta F_{a+i\in \va}$ after addition of a function 
node $a$ and all the edges $i$ around it, and the free
energy shift $\Delta F_{i}$ after addition of an edge $i=(ab)$.
These are given by:
\begin{eqnarray}
 e^{-\beta \Delta F_{a+i\in \va} } &=& e^{-\beta} + 
           \sum_{i \in a}  e^{\beta  h^{i \to a}}\, , \label{df3}\\
 e^{-\beta \Delta F_{i} } &=& 1 +  
            e^{\beta( h^{i \to a}+ h^{i \to b})} \, . \label{df4}
\end{eqnarray}
 The total free energy is then \cite{ MP99, YFW03}:
\begin{equation} 
F_G(\beta)=
\sum_a \Delta F_{a+i\in \va} - \sum_i \Delta F_{i}\, .
  \label{free_given}
\end{equation} 
This form of free energy is variational, i.e. the derivative 
$\frac{\partial (\beta F_G(\beta))}{\partial h^{i\to a}}$ vanishes if and only 
if the fields $h^{i\to a}$ satisfy (\ref{BP_T}). This allows to compute 
easily the internal energy
\begin{eqnarray}
   E_G(\beta) &=& 
    \sum_{a} \frac{  e^{-\beta} - \sum_{i\in a} h^{i\to a} 
    e^{\beta  h^{i\to a}} }
    { e^{-\beta} +\sum_{i\in a}  e^{\beta  h^{i\to a}}  } 
    + \sum_i \frac{(h^{i\to a}+ h^{i\to b}) e^{\beta(h^{i\to a}+ 
    h^{i\to b})}}{ 1 + e^{\beta(h^{i\to a}+ h^{i\to b})}}= \nonumber \\
    &=& N - 2 \sum_i \langle s_i \rangle   = \sum_a \frac{1}{1+\sum_{i \in a} 
    e^{\beta(1+h^{i\to a})}} .
    \label{energy_G}  
\end{eqnarray}
The second and third equalities have been derived using eq. (\ref{BP_T}). 
In the last term we can recognize the probability that a node $a$ is not 
matched. The entropy is then obtained as 
\begin{equation}
S_G(\beta)=\beta[E_G(\beta)-F_G(\beta)]\, .
\label{entropy_G}
\end{equation}

All the equations (\ref{cav_Z})-(\ref{entropy_G}) hold on a single large sparse
graph $G$. In section \ref{alg} we will describe how to use them to build
algorithms for counting and sampling the matchings on a given graph.  

We now study the typical instances in an   
ensemble of graphs. We denote the average over the ensemble by
$\mathbb{E}(\cdot)$. 
We assume that the random graph ensemble is given by a prescribed degree 
distribution ${\cal Q}(k)$.
Let us call ${\cal P}_\beta(h)$ the distribution of cavity fields over all the
edges of a large typical graph from the graph ensemble. It   satisfies the
following self-consistent  equation
\begin{equation}
 {\cal P}_\beta(h)= \sum_{k=1}^\infty \frac{k}{c} {\cal Q}(k)  
  \int \prod_{i=1}^{k-1} \left[ {\rm d} h^i {\cal P}_\beta(h^i) \right] 
  \delta{\left[h+ \frac{1}{\beta}
  \log{\left( e^{-\beta} +\sum_{i}  e^{\beta h^{i} } \right)}  \right]}.
  \label{pop_dyn}
\end{equation}
The term $k{\cal Q}(k)/c$ is the normalized degree distribution of the
function node $a$ when one picks up uniformly at random an edge $a-i$ from
the factor graph; 
$c= \sum_k k{\cal Q}(k)$ is the mean degree.  
This equation for distribution ${\cal P}_\beta(h)$ can be 
solved numerically by a technique of population dynamics \cite{MP99}.

The average of the free energy density is  then 
\begin{eqnarray}
  f(\beta) &=& \frac{{\mathbb{E}}[F_G(\beta)]}{N}  = -\frac{1}{\beta} 
    \sum_{k=0}^\infty {\cal Q}(k) 
    \int \prod_{i=1}^k \left[ {\rm d} h^i {\cal P}_\beta(h^i)\right] 
    \log{\left( e^{-\beta} +\sum_{i}  e^{\beta h^{i}} \right)} \nonumber \\
    &+& \frac{c}{2\beta} 
    \int {\rm d} h^1 \, {\rm d} h^2\, {\cal P}_\beta(h^1)\, 
    {\cal P}_\beta(h^2)\, \log{\left(1 + e^{\beta(h^{1}+ h^{2})}\right)}.
    \label{free_en}
\end{eqnarray}
This expression for the free energy  is in its variational form 
(see \cite{MP99}), i.e. the functional derivative 
$\frac{\delta f(\beta)}{\delta {\cal P}_\beta(h)}$ vanishes if and only if
 ${\cal P}_\beta$ satisfies
(\ref{pop_dyn}). The average energy density is then equal to:
\begin{eqnarray}
   \epsilon(\beta) &=&  \sum_{k=0}^\infty {\cal Q}(k) 
    \int \prod_{i=1}^k \left[ {\rm d} h^i {\cal P}_\beta(h^i)\right]
    \frac{  e^{-\beta} - \sum_{i} h^{i} e^{\beta  h^{i}} }
    { e^{-\beta} +\sum_{i}  e^{\beta  h^{i}}  } \nonumber \\
    &+& \frac{c}{2} 
    \int {\rm d} h^1 \, {\rm d} h^2\, {\cal P}_\beta(h^1)\, 
    {\cal P}_\beta(h^2)\, 
    \frac{(h^{1}+ h^{2}) e^{\beta(h^{1}+ h^{2})}}
    { 1 + e^{\beta(h^{1}+ h^{2})}  }.
    \label{energy}  
\end{eqnarray}
The average entropy density is 
\begin{equation}
s(\beta) = \beta [ \epsilon (\beta) - f(\beta) ]\, . \label{entropy}
\end{equation}

All our computations up to now rely on the only
assumption (the 'RS cavity assumption') that  the neighbors of a node in a 
cavity are  uncorrelated. A
necessary condition for the validity of this assumption is
that the following nonlinear (spin-glass) susceptibility be finite 
\cite{MR03, RBMM04}:
\begin{equation}                                              
    \chi_{SG}=\sum_i {\mathbb{E}}(\langle s_0 s_i \rangle^2_c)=    
    \sum_{d=0}^\infty \alpha^d \, {\mathbb{E}}(\langle s_0 s_d \rangle^2_c)  \ .
    \label{chi2}
\end{equation}                                             
Here $\langle s_0 s_i \rangle_c$ is the connected correlation
function between reference edge~$0$ and edge~$i$, $\alpha^d$ 
is the average number of vertices at distance $d$ from the edge $0$,
for general degree distribution $\alpha=\sum_{k=0}^\infty k(k+1) 
{\cal Q}(k+1)/c$.
The susceptibility is finite if and only if         
\begin{equation}                                              
   \overline \lambda_T = \lim_{d\to \infty}\alpha  \left[ 
   {\mathbb{E}}(\langle s_0 s_d \rangle^2_c)
   \right]^{\frac{1}{d}} < 1\, .             
   \label{stabT}
\end{equation}    
We will call $\overline \lambda_T$ the finite temperature stability parameter. 
A necessary condition for the RS cavity assumption to hold is that  
$\overline \lambda_T<1$.  

\begin{figure}[!ht]
  \begin{center}
   \resizebox{3cm}{!}{\includegraphics{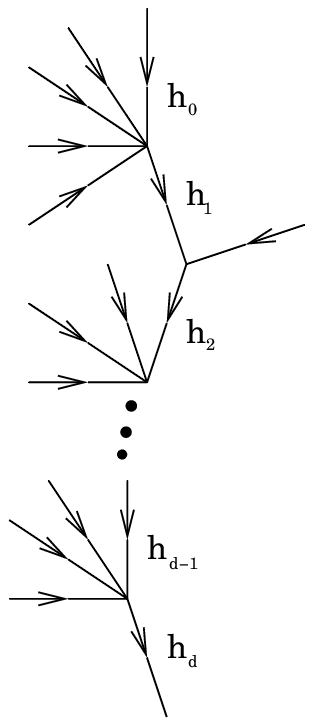}}
  \end{center}
   \caption{\label{chain} Chain of the cavity fields used to compute the 
   finite temperature stability parameter.}
\end{figure}
         
Using the fluctuation-dissipation relation when edge $i$ is 
at distance~$d$ from edge~$0$ we have for the correlation
function
\begin{equation}                                              
   {\mathbb{E}}(\langle s_0 s_d \rangle^2_c) = {\cal C} \,        
  {\mathbb{E}}  \left[\left( \frac{\partial 
    h_d } {\partial h_0}  \right)^2\right] = {\cal C} \,  {\mathbb{E}} 
    \left[  \prod_{i=1}^d \left( \frac{\partial h_{i} } 
    {\partial h_{i-1}} \right)^2\right] ,
    \label{fdt}
\end{equation}                           
where ${\cal C}$ is a $d$-independent constant. 
The field $h_i$ is according to (\ref{BP_T}) a function of 
$h_{i-1}$ and other fields $h_{i-1}^{(k)}$ incoming to $h_i$, 
see fig. \ref{chain}.
\begin{equation}
    h_{i} =  - \frac{1}{\beta} \log{\left[e^{-\beta}+ 
   e^{\beta h_{i-1}} + \sum_{k=1}^{p_{i-1}} e^{\beta h_{i-1}^{(k)}} \right]}\, . 
\end{equation}
The number $p_{i-1}$ of the incoming fields is chosen according to the 
probability distribution ${\cal Q}_2(p_{i-1})$ of the number of neighbors 
of a node given this node has already two other neighbors, 
${\cal Q}_2(k) = (k+2)(k+1) {\cal Q}(k+2)/ (\alpha c)$; in particular for Poisson
distributions ${\cal Q}_2={\cal Q} $. 
The values of the fields $h_{i-1}^{(k)}$ are chosen randomly from 
the distribution (\ref{pop_dyn}).

\subsection{Zero temperature limit}

The zero temperature limit ($\beta \to \infty$) corresponds to 
the ground state (maximum matching) of our system. 
Let us investigate the explicit behavior of that 
limit. 

Our numerical studies of (\ref{pop_dyn}) show that for large 
$\beta$ the cavity field distribution ${\cal P}_\beta(h)$ 
peaks around three different values $h\in \{0,\pm 1 \}$.
\begin{equation}
  {\cal P}(h) = p_1 \delta(h-1) + p_2 \delta(h+1) + 
   p_3 \delta(h)\, , 
   \label{dist_P}
\end{equation} 
where $p_1$, $p_2$ and $p_3$ are the weights (probabilities) 
of $h=1$, $-1$ and $0$. The cavity fields update (\ref{BP_T}) becomes 
\begin{equation} 
 h^{i\to a}= - \max_{j\in b - i}(-1, h^{j\to b})\, . 
  \label{BP_0}
\end{equation}

These equations may also be derived by working directly at zero 
temperature as in \cite{MP03}.
We define the cavity energy $E_{s_i}^{i\to a}$ as the ground state energy
of subgraph containing edge $i$ when constraint $a$
is absent (fig. \ref{fac_gr}) and edge $i$ takes value
$s_i$. The analog of (\ref{cav_Z}) is
\begin{equation}
     E_{s_i}^{i\to a} = \min_{\{s_j\}}\, \ind\left( s_i +\sum_{j\in b -i}
     s_j\le 1 \right)\left[ \sum_{j\in b -i}
    E_{s_j}^{j\to b} + (1 -s_i -\sum_{j\in b -i} s_j)\right].
    \label{cav_en}
\end{equation}
If one defines the cavity fields as 
\begin{equation}
      h^{i\to a} =  E_{0}^{i\to a} -  E_{1}^{i\to a}  
\label{hdef_physical}
\end{equation}
then (\ref{cav_en}) gives  back the cavity fields update (\ref{BP_0}).
The difference between cavity 
energies when $i$ is (is not) matched may be $\pm 1$ or $0$ and those are the 
three possible values of cavity fields.

Eq. (\ref{pop_dyn}), taken in the $\beta \to \infty$ limit, shows that
\begin{eqnarray}
      p_1 &=&\frac{1}{c} \sum_{k=0}^\infty (k+1) {\cal Q}(k+1) p_2^k\, , 
      \label{p1}\\
      p_2 &=& \frac{1}{c} \sum_{k=0}^\infty (k+1) {\cal Q}(k+1) [1-(1-p_1)^k]\, ,
      \label{p2} \\
      p_3 &=& \frac{1}{c} \sum_{k=0}^\infty (k+1){\cal Q}(k+1)[(1-p_1)^k-p_2^k]\, .
      \label{p3}
\end{eqnarray}
The possible solutions to these equations depend on the distribution 
${\cal Q}(k)$. 
\begin{itemize}
  \item[(a)]{There always exists a solution with 
        $p_3=0$, $p_1=1-p_2$.}
  \item[(b)]{For graphs without leaves, ${\cal Q}(1)=0$, there exists a 
        solution with $p_3=1$, $p_1=p_2=0$.}
  \item[(c)]{For graphs with leaves ${\cal Q}(1)>0$ a solution with 
        $0 < p_1, p_2, p_3 < 1$ exists if the mean degree is 
        sufficiently large.}
\end{itemize}
Let us stress at this point that our numerical solution of 
(\ref{pop_dyn}) for the cavity fields distribution at very small 
but nonzero temperatures corresponds to case $p_3>0$, (b) or (c). In other
word whereas at zero temperature there exist two mathematically possible 
solutions of (\ref{p1})-(\ref{p3}), at arbitrary small temperature 
only the one with $p_3>0$ exists. In the rest of this section 
we describe this ``small temperature'' solution. Case (a), which 
exists only at strictly zero temperature, and which forbids the 
cavity fields $h=0$, will be discussed in section \ref{1RSB}. 
 
Using (\ref{free_en}) the ground state energy, related to the size of the 
maximum matching, is
\begin{eqnarray}
    \epsilon_0 
    = {\cal Q}(0) + \sum_{k=1}^\infty {\cal Q}(k) [ p_2^k + (1-p_1)^k -1 ] +
    c p_1(1-p_2)\, .  \label{e_0_gen}
\end{eqnarray}
If we consider solution (b) for $p_1$, $p_2$, $p_3$ for graphs with 
no leaves,  $Q(1)=0$, then the ground state energy is $\epsilon_0= Q(0)$, 
i.e. asymptotically all the non-isolated vertices are matched. 
In other words, in an ensemble of random graphs with minimal degree 
$2$ (e.g. regular graphs) almost every graph has an almost perfect matching. 
This is in agreement with the result of Karp and Sipser \cite{KS81} and also 
with a stronger result of Frieze and Pittel \cite{FP03}, who also found 
the (small) number of vertices which cannot be matched.  

To compute the average ground state entropy we need to expand the 
free energy at low temperatures $f(\beta \to \infty)=
e_0-s_0/\beta+O(1/\beta^2)$. This requires to study the 
``evanescent'' parts of the cavity fields \cite{BMW00}, i.e. the leading 
corrections to their value at $\beta=\infty$. 
Numerically we have observed that at $\beta \gg 1$
the three delta peaks
(\ref{dist_P}) keep their weights ($p_1, p_2, p_3$) and spread as
\begin{eqnarray}
      h&=& 1+ \frac{\log{\nu}}{\beta} \quad   
      {\rm for\  the\  peak\  around} \quad  h=1\, ,
          \label{h1} \\
      h&=& -1+\frac{\log{\mu}}{\beta} \quad   
      {\rm for\  the\  peak\  around} \quad  h=-1\, ,
          \label{h2} \\
      h&=& \frac{\log{\gamma}}{\beta} \quad   
      {\rm for\  the\  peak\  around} 
         \quad  h= 0\, .
          \label{h3}
\end{eqnarray}
From (\ref{pop_dyn}) we derive self-consistently the 
distributions ${\cal A}$ of the evanescent cavity fields 
$\nu, \mu, \gamma$
\begin{eqnarray}
      {\cal A}_1(\nu) &=& \sum_{k=0}^\infty {\cal C}_1(k) 
              \int \prod_{i=1}^k \left[  {\rm d}\mu_i {\cal A}_2(\mu_i) \right] 
              \delta\left(\nu - \frac{1}{1+\sum_i \mu_i}\right)    ,
              \label{a1_gen} \\
      {\cal A}_2(\mu) &=& \sum_{k=1}^\infty  {\cal C}_2(k)
              \int \prod_{i=1}^k \left[  {\rm d}\nu_i {\cal A}_1(\nu_i) \right] 
              \delta\left(\mu - \frac{1}{\sum_i \nu_i}\right) ,
               \label{a2_gen} \\
     {\cal A}_3(\gamma)&=&\sum_{k=1}^\infty {\cal C}_3(k)
              \int \prod_{i=1}^k \left[  {\rm d}\gamma_i {\cal A}_3(\gamma_i) 
              \right] 
              \delta\left(\gamma - \frac{1}{\sum_i \gamma_i}\right),
              \label{a3_gen}
\end{eqnarray}
where the combinatorial factors ${\cal C}_1, {\cal C}_2, {\cal C}_3$  
are given by
${\cal C}_1(k)=  
\frac{(k+1) p_2^k  {\cal Q}(k+1)}{p_1 c}$, ${\cal C}_2(k)= \frac{p_1^k}{k!}
\sum_{m=k}^\infty \frac{(1-p_1)^{m-k}{\cal Q}(m+1)(m+1)!}{(m-k)!p_2 c}$, 
${\cal C}_3(k) = \frac{p_3^k}{k!} \sum_{m=k}^\infty \frac{p_2^{m-k}
{\cal Q}(m+1)(m+1)!}{(m-k)!p_3 c}$.
Using eqs. (\ref{h1})-(\ref{a3_gen}) we expand the free energy 
to order $1/\beta$ and get the ground state entropy of maximum 
matchings
\begin{eqnarray}
   s_0 &=& \sum_{k=1}^\infty \frac{p_3^k}{k!}\sum_{m=k}^\infty 
       \frac{p_2^{m-k}{\cal Q}(m)m!}{(m-k)!}\, 
       \overline{\log{\left(\sum_{i=1}^k \gamma_i\right)}}
       -cp_1p_3\,  \overline{\log{\gamma}} - 
       \frac{cp_3^2}{2}\, \overline{\log{(1+\gamma_1\gamma_2)}} \nonumber \\
       &\, & -cp_1(p_1+p_3)\, \overline{\log{\nu}}+
       \sum_{k=1}^\infty \frac{p_1^k}{k!}\sum_{m=k}^\infty 
       \frac{(1-p_1)^{m-k}{\cal Q}(m)m!}{(m-k)!}\, 
       \overline{\log{\left(\sum_{i=1}^k \nu_i\right)}} \nonumber  \\
       &\, &+\sum_{k=0}^\infty {\cal Q}(k)p_2^k \, 
       \overline{\log{\left(1+\sum_{i=1}^k \mu_i\right)}}
       -cp_1p_2\, \overline{\log{(1+\mu\nu)}}\, ,
       \label{ent_gen}
\end{eqnarray}
where the overlines denote expectations over independent random 
variables with distribution ${\cal A}_1$ (for $\nu$-variables), ${\cal A}_2$ 
(for $\mu$-variables), ${\cal A}_3$ (for $\gamma$-variables).

To conclude this section we describe the zero temperature version 
of the stability analysis for the cavity assumption. What follows
is equivalent to the stability analysis of 
the replica symmetric assumption with respect to replica symmetry 
breaking for discrete sets of cavity fields \cite{TKC05}. 
Here we will describe this stability analysis in an intuitive 
way as a spread of changes in the cavity fields update that is 
analogous to what is called {\it bug proliferation} in the 
context of the stability of the one step replica symmetry breaking ansatz 
\cite{MR03, RBMM04, MMZ05}.
  
Consider a node $b$ with $k+1$ neighbors, fig. \ref{fac_gr}. 
Choose one incoming cavity field $h^{j\to b}$ and one 
outgoing field $h^{i\to a}$. 
Now consider probability  $P_k(\alpha_o \to \gamma_o |\alpha_i \to \gamma_i  )$
that the value of the outgoing field change 
from $\alpha_o \in \{\pm 1,0 \}$ 
to $\gamma_o\in \{\pm 1,0 \}$ providing the incoming one has been changed 
from $\alpha_i\in \{\pm 1,0 \}$ to $\gamma_i\in \{\pm 1,0 \}$. 
More precisely $P$ is probability of having a set of other $k-1$ incoming
fields such that it causes the change $\alpha_o \to \gamma_o$ given that 
the change $\alpha_i \to \gamma_i$ happened. 
There are eight different combinations of cavity fields such 
that $P_k(\alpha_o \to \gamma_o |\alpha_i \to \gamma_i  )$ is nonzero
\begin{eqnarray}
  P_k(1\to -1|-1\to 1) = P_k(-1\to 1|1\to -1)&=& p_2^{k-1}, \\
  P_k(1\to  0|-1\to 0) =P_k(0\to 1| 0\to -1) &=& p_2^{k-1}, \\
  P_k(-1\to 0|1\to 0)= P_k(0\to -1| 0\to 1)& =& (p_2+p_3)^{k-1}, \\
  P_k(-1\to 0|1\to -1)= P_k(0\to -1| -1\to 1)  &=& (p_2+p_3)^{k-1}- p_2^{k-1}.
\end{eqnarray}
In the first step we change a cavity field from $\alpha_i$ to $\gamma_i$.
Average probability of change $\alpha_o$ to $\gamma_o$ that follows is 
\begin{equation}
    \overline{P}(\alpha_o \to \gamma_o |\alpha_i \to \gamma_i ) 
    = \sum_{k=0}^\infty {\cal Q}_2(k) 
    P_{k+1}(\alpha_o \to \gamma_o |\alpha_i \to \gamma_i).   
\end{equation}

The most important change is given by the largest 
eigenvalue $\overline \lambda_{\rm max}$ of the matrix $\overline{P}$. In analogy
with (\ref{chi2}) we are interested in stability parameter 
$\overline \lambda_{0}= \alpha \overline \lambda_{\rm max}$ 
\begin{equation}
  \overline \lambda_{0} =\alpha  \sqrt{\sum_{k=0}^\infty   
  {\cal Q}_2(k)\,  p_2^k} \, \sqrt{ \sum_{k=0}^\infty   
  {\cal Q}_2(k) \, (p_2+p_3)^k  }    \, . 
  \label{lambda_0_av}
\end{equation}
If $\overline \lambda_0>1$ the total number of changes after many steps 
will diverge and we cannot hope the cavity assumption to be valid. 
On the other hand if $\overline \lambda_0<1$ then the first change 
will not spread very far and the RS assumption is locally stable. 

Note also that $\overline \lambda_0$ and $\overline \lambda_{T\to 0}$ are not 
equal, because they count different quantities. But we expect that their
position relative ti the threshold value $1$ is the same. In other words 
both of them correctly describe the stability at zero temperature. 
The advantage of  $\overline \lambda_0$ is that it is far more easy to 
compute than the $d \to \infty$ limit of $\overline \lambda_{T,d}$ 
(\ref{stabT})-(\ref{fdt}).

\subsection{Algorithms following from the cavity method}
\label{alg}

\subsubsection{Zero temperature message passing and leaf removal}
\label{core}

The zero temperature limit of the cavity fields update 
(\ref{BP_0}) can be seen as a message passing 
(warning propagation) algorithm. Interpretation of the three 
different cavity fields is: $h=1$ means {\it ``I want you to match me''}, 
$h=-1$ means {\it ``I want you not to match me''}, $h=0$ means 
{\it ``No preferences, do what you want''}. 
The interpretation of the cavity fields update (\ref{BP_0}) is: 
\begin{itemize} 
    \item{If one or more of my neighbors wants me to match them, 
         I match one of them, and I send: {\it do not match me}. }
    \item{If none of my neighbors wants me to match it, and at least 
         one of them does not have any preferences 
         I send: {\it no preferences, do what you want}.}
    \item{If all of my neighbors are saying do not match me, 
         or if I have no neighbors, I send: {\it match me}. }
\end{itemize}

This message passing procedure  
starting from all the $h=0$ is equivalent to the step (1) of the Karp and 
Sipser's leaf removal procedure in the following sense: 
Run the message passing until you find a fixed point. Then the edges 
where a message $h=1$ or $h=-1$ is sent from at least one side are 
exactly those edges which have been matched or removed in the leaf 
removal procedure. Consequently the 
edges in the core are those which have $0$ sent from both sides. Using eq. 
(\ref{BP_T}) at a very small temperature we can use arbitrary initial 
conditions. 

\subsubsection{Uniform sampling}

Solving the BP equations (\ref{BP_T}) on a given graph $G$  by iterations 
allows to sample typical matchings from Boltzmann's distribution 
(\ref{Boltzmann}) at inverse temperature $\beta$ 
(i.e. matchings of size $[1-E_G(\beta)]/2$).
 
Such a sampling can be done as follows: one chooses a variable 
node $i$, computes $\langle s_i\rangle$ from (\ref{marginal}), 
and generates the value of $s_i$ as $s_i=1$ with probability 
$\langle s_i\rangle$ , and $s_i=0$ with probability $1-\langle s_i\rangle$.
Once $s_i$ has been fixed, this imposes that all the fields 
$h^{i\to a}$ (for all function nodes $a$ connected to $i$) 
are equal either to $+\infty$ (if $s_i=1$) or to $-\infty$ (if $s_i=0$).
One runs again the BP equations, with these extra constraints, 
and iterate this procedure.

\subsubsection{Counting matchings on a single graph}
\label{counting}

Our results may be also used to estimate the size (\ref{energy}) 
and number of matchings (\ref{entropy}) on arbitrary spare large graph 
$G$. The size of the maximum matching is obtained from the zero temperature 
limit of (\ref{free_given}) or (\ref{energy_G}), this is not very 
interesting since the solution to this problem is well known, e.g. 
\cite{MV80}. An algorithm to compute approximately the number of 
matchings of given size is more interesting.
 
\vspace{0.5cm}

{\tt

\noindent INPUT: The graph, the inverse temperature $\beta$, a maximum 
number of iterations $t_{max}$. \\
\noindent OUTPUT: The entropy of matchings $S_G=\log{\cal N}_G$ of size 
$(1-E_G)/2$. If at the end $E_G=-1$ the procedure failed to converge.
       
\begin{itemize}
  \item[1.]{Initialize all the cavity fields $h^{i\to a}$ to some random value.}
  \item[2.]{Iterate belief propagation equations (\ref{BP_T}) until they 
    converge, i.e. the values of the cavity fields do not change 
    anymore, or until the 
    number of iterations exceeds $t_{max}$.} 
  \item[3.]{$E_G=-1$. If the number of steps is $>t_{max}$ STOP. 
    Else: compute the energy $E_G$ and the free energy $F_G$ of matchings 
    corresponding to $\beta$ according to (\ref{energy_G}) and 
    (\ref{free_given}), compute the entropy $S_G=\beta[E_G-F_G]$.} 
\end{itemize}
 
}

In order to compute the total number of matchings one needs to take 
the $\beta \to 0$ limit. To compute the number of maximal matchings 
one takes $\beta \to \infty$. In both cases the algorithm can be 
rewritten and simplified. 

Note also that the complexity of this algorithm is only linear in 
number of edges. The convergence and correctness in the highest 
order ($(\log{\cal N}_G)/N$) for large sparse graphs or trees is 
expected for the same reasons as the correctness of results 
of cavity method for the ensembles of $r$-regular and ER random graphs. 

On small or loopy graphs the cavity fields update does not have 
to converge or its fixed point may depend on the initial conditions. 
However, it would be interesting to apply it to ``real world''
graphs as in \cite{loops}, or to compare the results with those 
of existing methods \cite{JS97, FK04, Chien04}. 

\section{Application to random graph ensembles}
\label{results}

We compute and discuss the results for two random graph 
ensembles, the $r$-regular and ER graphs.

\subsection{Random regular graphs}
\label{res_reg}

For $r$-regular regular graphs (${\cal Q}(k)=\delta_{kr}$) the results 
are particularly simple. All the vertices are equivalent in the 
cavity method. It means that the solution of (\ref{pop_dyn}) is
${\cal P}_\beta(h)=\delta(h-h_r)$, where $h_r$ is the solution of  
\begin{equation}
           h_r = -\frac{1}{\beta} \log{[e^{-\beta}+(r-1)e^{\beta{h_r}}]}\ ,
       \label{h_reg}
\end{equation}
given by:
\begin{equation}
h_r=\frac{1}{\beta}\log\left[\frac{\sqrt{4(r-1)+e^{-2\beta}}-e^{-\beta}}
{2(r-1)}\right]\, .
\end{equation}
The free energy density (\ref{free_en}) simplifies to
\begin{equation}
           f = -\frac{1}{\beta} \log{[e^{-\beta}+r e^{\beta{h_r}}]} +
             \frac{r}{2\beta} \log{[1+e^{2\beta{h_r}}]}\, .
          \label{free_reg}
\end{equation}
The energy density (\ref{energy}) reads
\begin{equation}
          \epsilon = \frac{e^{-\beta}-r h_r e^{\beta h_r}}
              {[e^{-\beta}+r e^{\beta{h_r}}]} +
             \frac{ r h_r e^{2\beta h_r}}{[1+e^{2\beta{h_r}}]}\, .
           \label{energy_reg}
\end{equation}
The entropy, related to the number of matchings, is computed using 
(\ref{entropy}). With a bit 
of algebra one finds that the quenched entropy 
as a function of energy is equal to the annealed result of 
eq. (\ref{ent_reg_ann}), 
where $\epsilon =1-x$. This result is compatible with, but 
slightly stronger than the Theorem 1 of Bollob\'as and McKay 
\cite{BM86}.

The matching problem is equivalent to the physical problem of dimers on 
a graph. There it is natural to compute the density of dimers $rp$ as a 
function of $\beta$ (which is half of the chemical potential in the 
context of dimers). For $r$-regular graphs we find that  this 
``equation of state'' is
\begin{equation}
       e^{2\beta} = \frac{p(1-p)}{(1-rp)^2}\, .
\end{equation}
This result has already been obtained in \cite{HC04} for the dimer problem on 
Bethe lattice.

For $r \ge 2$ in the zero temperature limit  
one has  $h_r=0$, which corresponds to the solution (b) of 
(\ref{p1})-(\ref{p3}). The ground state energy density is then $\epsilon_0=0$, 
this means that asymptotically almost all the vertices may be matched 
for almost every $r$-regular graph. This agrees with the stronger mathematical 
result that for $r\ge 3$ there exists a perfect matching almost surely
in graphs with even number of vertices. 
From (\ref{ent_gen}) one finds that the ground state entropy density
 $\lim_{\beta \to \infty} s(\beta)$ is 
\begin{equation}
       s_0=[(r-1)\log{(r-1)}-(r-2)\log{r}]/2\, , \label{ent_reg_0}   
\end{equation}
in agreement with the annealed result (\ref{ent_reg}) of Bollob\'as and McKay.

The stability parameter $\overline \lambda_T$ (\ref{stabT}) for $r$-regular
graphs is 
\begin{equation}
     \overline \lambda_T= (r-1) \left( 
     \frac{\partial h_1}{\partial h_0}\Big|_{h_1=h_0=h_r} \right)^2 =
     (r-1) \left[\frac{\sqrt{4(r-1)+e^{-2\beta}}-e^{-\beta}}{2(r-1)}\right]^4\, .
\end{equation}
We see that $\overline \lambda_T<1$ for all finite temperatures and $r\ge 2$. 
In the zero temperature limit $\overline \lambda_{T\to 0}=1/(r-1)$. 
The zero temperature stability parameter (\ref{lambda_0_av}) for $r$-regular
graphs is $\overline \lambda_0 = 0$ for $r > 2$, and $\overline 
\lambda_0 = 1$ for $r = 2$. This agrees qualitatively with the behavior of
$\overline \lambda_{T\to 0}$. It is worth noticing that also the 
ferromagnetic stability parameter,
defined as $(r-1) \left(\frac{\partial h_1}{\partial h_0}
\Big|_{h_1=h_0=h_r} \right) $, is at finite temperature smaller than 
1 when $r\ge 2$.
It should be possible to use this result in order to show that the 
matching properties on the root of a large tree are completely 
independent of boundary conditions, which
could then allow for a rigorous proof of our results following the lines of 
Bandyopadhyay and Gamarnik \cite{BG05}.

In order to exclude the possibility of a discontinuous transition 
towards a phase with
broken replica symmetry, which we cannot see by analyzing the stability, 
we wrote the 1RSB \cite{MP99} equations for the $r$-regular graph. 
We have seen 
clearly numerically that their solution reduces to the replica symmetric 
one. So all the evidence suggests that the RS cavity assumption should 
be valid for $r$-regular graphs and so we expect our result for the quenched 
entropy to be exact.

\subsection{ER graphs}
\label{res_ER}

\begin{figure}[!ht]
  \begin{center}
   \resizebox{10cm}{!}{\rotatebox{0}{\includegraphics{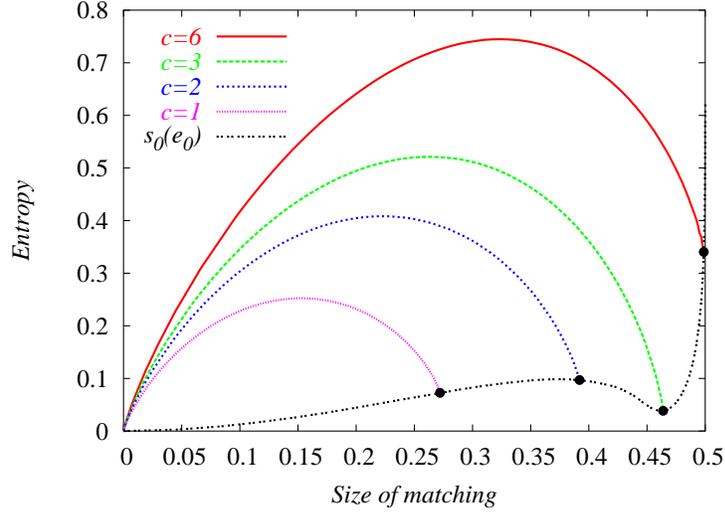}}}
  \end{center}
   \caption{\label{s_e} Entropy density $s(m)=\log{{\cal N}(m)}/N$ as a 
      function of relative size of the matching $m=|M|/N=(1-\epsilon)/2N$ 
      for ER random graphs with mean degrees $c=1, 2, 3, 6$.
      The lower curve is the ground state entropy density 
      $s_0(\epsilon_0)$ for all mean degrees. 
      The curves are obtained by solving eqs. (\ref{pop_dyn})-(\ref{entropy})
      with a population dynamics, using a population of sizes  
      $N=2\cdot 10^4$ to $2\cdot 10^5$ and the number of iterations 
      $t_m=10000$.}
\end{figure}

For the ER random graphs, with degree distribution 
${\cal Q}(k)= e^{-c}c^k/k!$, 
we have solved numerically the equation (\ref{pop_dyn}) by the 
population dynamics method. Using (\ref{free_en})-(\ref{entropy})
we have then computed the energy density $\epsilon(\beta)$ 
(related to the size of the matching as $|M|=(1-\epsilon)N/2$) 
and the entropy density $s(\beta)$ for values of 
$\beta \in (-\infty, \infty )$. In fig. \ref{s_e} we show
the entropy versus the size of matching for mean degrees $c=1,2,3$ and $6$. 

The maxima of the curves in fig. \ref{s_e} give the entropy of 
all the possible matchings, regardless of their sizes. The lower right ends 
of the curves gives the ground state energy 
(the size of the maximal matching) and the ground state entropy 
(number of maximal matchings). They are computed using by the following 
direct zero temperature method. 

The zero temperature equations (\ref{p1})-(\ref{p3}) for the Poissonian 
distribution become
\begin{eqnarray}
      p_1 &=& e^{-c(1-p_2)}\, , \qquad
      p_2 = 1 - e^{-cp_1}\, ,\qquad p_3=1-p_1-p_2\, . \label{p12} 
\end{eqnarray}
Analyzing these equations we can see that for $c\le e$ there exists only 
the solution (a), with $p_1+p_2=1$ and $p_3=0$. For $c > e$ there exists 
a second solution (c) with $p_1+p_2<1$ and $1>p_3>0$.   
From the population dynamics solution of eq. (\ref{pop_dyn}) at very small 
temperatures for $c>e$ we found that the solution (c) with $p_1+p_2<1$ is 
the proper zero temperature limit for $c > e$. 

The ground state energy (\ref{e_0_gen}) then reads
\begin{equation}
   \epsilon_0 = 1-2\frac{|M^*|}{N} = -p_2+p_1 +cp_1 -cp_1p_2\, . \label{e_0_1}
\end{equation} 
This is the exact result of Karp and 
Sipser \cite{KS81}, Theorem 2.     

The ground state entropy for ER graphs is computed using population 
dynamics equations (\ref{a1_gen})-(\ref{a3_gen}) with combinatorial factors
\begin{equation}
     {\cal C}_1(k)= \frac{e^{-cp_2}(cp_2)^k}{k!}\, , \quad 
     {\cal C}_2(k)= \frac{e^{-cp_1}(cp_1)^k}{(1-e^{-cp_1})k!}\, , \quad 
     {\cal C}_3(k)= \frac{e^{-cp_3}(cp_3)^k}{(1-e^{-cp_3})k!}\, .
    \label{coef}
\end{equation}
Factors ${\cal C}_i(k)$ are properly normalized Poissonian distribution 
with mean equal to concentration of corresponding cavity fields.  
The ground state entropy (\ref{ent_gen}) finally simplifies to
\begin{eqnarray}
   s_0 &=& -(1+cp_1)p_3\, \overline{\log{\gamma}} - 
       \frac{cp_3^2}{2}\, \overline{\log{(1+\gamma_1\gamma_2)}} \nonumber \\
       &\, &-p_2\, \overline{\log{\mu}} -p_1(1+cp_1+cp_3)\, \overline{\log{\nu}}
       -cp_1p_2\, \overline{\log{(1+\mu\nu)}}\, .
       \label{ent_poiss}
\end{eqnarray}

We call the first two terms in eq. (\ref{ent_poiss}) the {\it core entropy}
$s_c$, the averages (denoted by overlines) are over the distribution 
(\ref{a3_gen}). The rest 
(last three terms) we call the {\it non-core entropy} $s_{nc}$, the averages
are over the distributions (\ref{a1_gen}) and (\ref{a2_gen}). 
The reason for these names is the following. 
Since we know the size and degree distribution on the core 
(\ref{core_deg}) we can use eq. (\ref{ent_gen}) directly only 
for the core and indeed we will obtain the first two terms of 
eq. (\ref{ent_poiss}), the core entropy. The rest is the entropy
corresponding to the choice of the matching in the non-core part 
of the graph.

\begin{figure}[!ht]
  \begin{center}
   \resizebox{11cm}{!}{\rotatebox{0}{\includegraphics{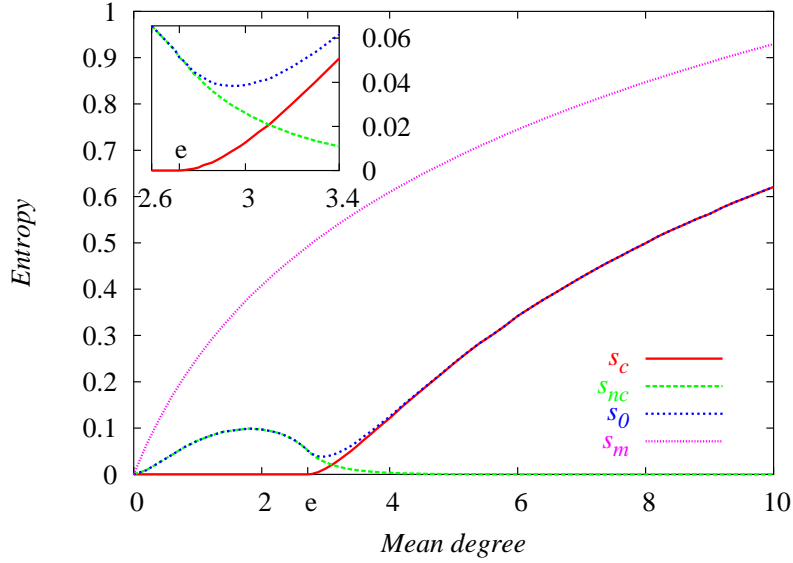}}}
  \end{center}
   \caption{\label{s_c} The ground state entropy density $s_0$ (giving 
      leading exponential behaviour of the 
      number of maximum matchings) and the full entropy $s_m$ (giving 
      leading exponential behaviour of 
      the number of all possible matchings) as a function of mean
      degree $c$ in ER random graphs. The detail is in inset.   
      The ground state entropy is the sum of $s_c$, the contribution of 
      the core,
      and $s_{nc}$, the contribution or the parts of graph removed in 
      the leaf removal procedure. We see that $s_c>0$ only 
      for $c>e$, because the core 
      covers a finite fraction of vertices only when $c>e$.}
\end{figure}

Fig. \ref{s_c} shows the core and non-core entropies of the maximum 
matchings and their sum as a function of the average degree $c$.
The fourth (upper) line in fig. \ref{s_c} is the total entropy of all 
matchings. 

\begin{figure}[!ht]
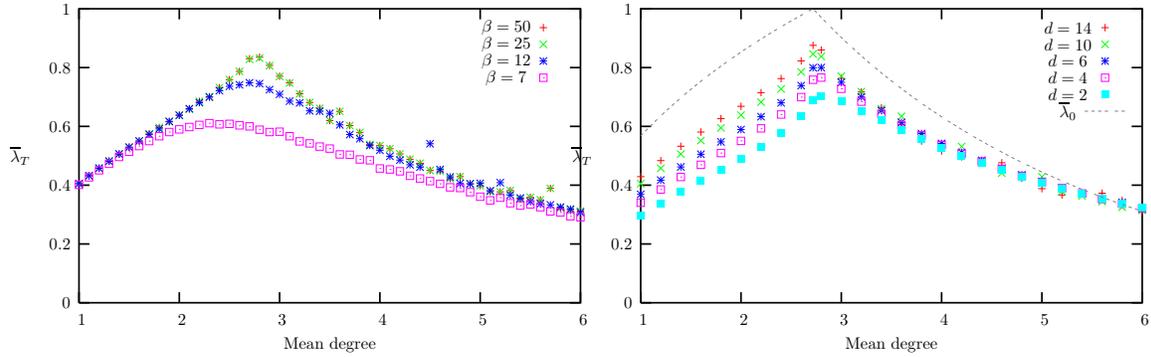

\begin{minipage}{0.49\linewidth}
   \resizebox{8cm}{!}{\rotatebox{0}{\input{lambda_c2.tex}}}
\end{minipage}  
\begin{minipage}{0.49\linewidth}
   \resizebox{8cm}{!}{\rotatebox{0}{\input{stab_w_dep2.tex}}}
\end{minipage}  
   \caption{\label{stab} On the left the finite temperature stability 
   parameter $\overline \lambda_T$ (\ref{stab_poiss}) for distance $d=10$ as 
   a function of mean degree. The upper curve corresponds to the 
   smallest temperature ($\beta=50$). Data were 
   obtained for size of population $N=40000$ and time $t=40000$.
   On the right the dependence of $\overline \lambda_T$ for temperature 
   $\beta=50$ on the mean degree and distance $d$. 
   We can see that $\overline \lambda_T$ 
   is growing slightly with $d$ in the regime $c<e$. 
   For larger $d$ we would need very big populations go 
   obtain reliable data. The continuous line depict the zero temperature
   stability parameter $\overline \lambda_0$, eq. (\ref{lambda_0_Poiss}). }
\end{figure}

The finite temperature stability parameter $\overline \lambda_T$ 
(\ref{stabT}) for ER random graphs is 
\begin{equation}
     \overline \lambda_T = c  \left[ {\mathbb{E}}(\langle n_0 n_d \rangle^2_c)
     \right]^{\frac{1}{d}}.
     \label{stab_poiss}
\end{equation}
We have to compute it numerically as described on fig. \ref{chain} 
and eqs. (\ref{stabT})-(\ref{fdt}). As for the $r$-regular graphs we find 
that $\overline \lambda_T$ grows as temperature decreases, 
see on the left on fig. \ref{stab}. 
So we may analyze only the zero temperature limit, and if that one is 
stable, then also the finite temperature is stable. On the right on 
fig. \ref{stab} we can also see the dependence 
of $\overline \lambda_T$ on the distance $d$.
Although we are not able to compute precisely its $d\to \infty$ limit, 
all the evidences speak for the fact that even for 
$d\to \infty$ the stability parameter $\overline\lambda_T$ never exceeds $1$. 

To check this, we look directly at the zero temperature stability parameter 
$\overline \lambda_0$ (\ref{lambda_0_av}) which for ER graph reads
\begin{equation}
  \overline \lambda_0 = cp_1 \sqrt{1+\frac{p_3}{p_1}}.
  \label{lambda_0_Poiss} 
\end{equation}
Its value is also depicted in fig. \ref{stab}. We can see that 
$\overline \lambda_0<1$ (stable) for all mean degrees except
$c=e$ where $\overline \lambda_0=1$ (marginally stable).
Supported by the numerical data in fig. \ref{stab} we 
expect that in the $d\to \infty$ limit the $\overline \lambda_T$
would behave in the qualitatively same way. 

From this analysis it is reasonable to conjecture that in
ER random graphs the replica symmetric cavity 
assumption is correct and all our results, in particular for the entropy, 
are exact. Another strong argument in favour of the validity of 
replica symmetry at any finite temperature will be given in section \ref{1RSB}.

\section{Ergodicity breaking at zero temperature}
\label{1RSB}
The size  of the maximum matchings in
ER graphs was studied recently by Zhou and Ou-Yang (Z-O) \cite{ZOY03}, using
 the cavity method directly at zero temperature \cite{MP03},
with a one step RSB solution. In this section we discuss the difference between
their approach and ours, in particular as far as RSB effects are concerned.
We keep to ER random graphs.

One should first emphasize that both approaches give the same result for the
size of the largest matching in ER graphs, and this result also agrees with
the rigorous value of Karp and Sipser. Our formalism is more general in two
respects. 1) we can work at finite temperature, which gives access to the full
distribution of the number of matchings versus their size. 2) We study the
limit of zero temperature ($\beta \to \infty$) keeping the leading corrections
of order $1/\beta$ in the fields, see (\ref{h1}-\ref{h3}); this allows to
study the entropy of maximal matchings.

The issue of RSB at zero temperature, which is present in the Z-O approach,
and absent in ours, is a somewhat subtle one. We shall just present a few
arguments of explanation.

First one should note that one does not expect ergodicity to be broken at any
finite temperature in this problem. We have not tried to write a formal proof
of this statement, but a first strong argument comes from the fact that the
energy barriers are finite. Let us define as {\it a step} the fact of removing
(adding) an edge from (to) a matching so that the new configuration is still a
matching. By adding an edge to a matching we lower the energy by $2$, whereas
by removing an edge we increase the energy by $2$. Using these steps one may
go from any matching $M$ to any other matching $M'$. Furthermore, if $|M'|\ge
|M|$, one can choose the steps in such a way that at every step the energy is
not higher than $E_M+4$. In other words energy barriers in the matching are at
most 4. This argument suggests that there should not be ergodicity breaking at
finite temperature, provided there are no diverging entropic barriers. Another
indication of ergodicity comes from the rapid mixing results of the
Monte Carlo procedure in the related problem of sampling perfect matchings in
bipartite graphs \cite{JSV04}.

Let us now focus on the zero temperature approach. The finite 
energy barriers between almost perfect matchings on the core become
effectively infinite at zero temperature so the breaking of ergodicity
cannot be excluded. The RS cavity method gives
the equations (\ref{BP_0}) for the cavity fields, easily derived from
(\ref{cav_en}). The more subtle issue is the support of the distribution of
cavity fields. Because a field $h^{i\to a}$ is defined as the difference
(\ref{hdef_physical}) of the ground state energies conditioned to $i$ being
absent/present in the matching, it is clear that $h^{i\to a}\in\{-1,0,1\}$,
and in ER graphs with $c>e$ one should thus choose between the solutions (a)
and (c) of eqs. (\ref{p1})-(\ref{p3}). If one considers the equations
(\ref{BP_0}) on a graph which is a tree, one finds that actually on all edges
$h_{i\to a}\in\{-1,1\}$. Because ER graphs are locally tree-like (seen from a
randomly chosen point, the subgraph of its environment up to a fixed distance
$d$ is a tree with probability one in the large $N$ limit), it is tempting to
restrict the cavity fields to $\pm 1$ values. This is what is done in Z-O.
Then the RS solution, for any value of the average degree $c$, is necessarily
solution (a). This solution is unstable towards 1RSB at $c>e$, which forces
one to study the 1RSB solution in this regime, as was done in Z-O. 

The 1RSB solution for the maximal matchings is able to nicely reconstruct the
information that is contained in the $h=0$ fields of our RS solution with
support on $\{-1,0,1\}$ as follows: Let us consider an edge $i\to a$ which
should pass $h^{i\to a}=0$ in our formalism. In the 1RSB formalism it passes
a message which is a probability distribution on the space of cavity fields,
with support $\{-1,1\}$, of the form $\alpha \delta_{h,-1}+ (1-\alpha)
\delta_{h,1}$; the distribution of $\alpha$ is related to our distribution
${\cal A}_3$ (\ref{a3_gen}). Consequently, the complexity computed by 
Z-O is equal to the
complexity of the core. This means that different almost perfect matchings 
on the core form the different states, each state containing only one of 
them (similarly as in the XOR-SAT problem \cite{MRZ03, CDMM02}, or in the 
multi-index matching \cite{MMR05}). It is interesting to notice that,
through the restriction of cavity fields to $h\in\{-1,1\}$, the Z-O method at
the RS level completely neglects loops, the effect of which is recovered only
at the 1RSB level. Conversely, the inclusion of the value $0$ in our cavity
fields allows to take into account loops directly, in which case RSB is not
needed.

This physical interpretation of the Z-O 1RSB solution is confirmed by its
stability analysis:  Using notations of \cite{MR03} one should
compute the type I instability (interpreted as state aggregation) of the 1RSB
solution. Type II instability (interpreted as division of states) is
irrelevant here, because each state corresponds to a single almost perfect 
matching on the core and cannot divide further. We have
found that the 1RSB solution is stable, but only if one considers maximal
matchings ($y\to\infty$ in the Z-O notation): any departure from this limit
mixes the various configurations and restores ergodicity, as expected.

\section{Conclusion and discussion}

We have argued that the replica symmetric cavity solution is exact
for counting matchings on random graphs. We have computed the size and 
quenched (typical) entropy in two random graph ensembles. For $r$-regular
graphs we have shown that the quenched entropy 
of matchings of a given size agrees with the annealed one. For the 
Erd\"os-R\'enyi random graphs we have shown how our method reproduces result 
of Karp and Sipser for the size of maximum matching, and we computed 
the quenched entropy of matchings of a given size, fig. \ref{s_e}.

Our method provides an algorithm for counting and uniform sampling of 
matchings on a given sparse graph, which should give the exact entropy for 
graphs with a girth that diverges in the large size limit. It would be very 
interesting to apply it to ``real world'' graphs, e.g. the internet, 
as in \cite{loops}. Also its systematic study on graphs with smaller girth 
and comparison with existing approximative methods \cite{JS97, FK04, Chien04} 
could reveal some interesting properties. 

There are two obvious generalizations  of the matching problem 
where we expect that our method could be used straightforwardly.
One is the matching with 
weights on edges (preferences to be matched) which is a  dimer model on 
random graphs with quenched disorder. 
Another generalization is that instead of allowing a vertex to have 
none or one ($k=1$) matched edge around itself, we could allow none or $k>1$
edges around a vertex to be matched. Then $k=2$ would mean we are 
interested in sets of loops, a model that has been studied recently 
in \cite{loops, loops2}. The case $k>2$, corresponding to k-regular subgraphs,
is being studied by \cite{core}. 

We hope that the replica 
symmetric nature of matching on random graphs should allow to turn 
all our result into rigorous theorems. In this respect there are two directions
which look particularly promising. One is to generalize the local weak 
convergence approach of \cite{BG05} in order to turn our results into 
rigorous theorems when $\beta$ is small enough and/or $c$
is far enough from $e$ (for ER graphs). The second one is to use Guerra's 
interpolation method \cite{Guerra03, FL03, FLT03} in order
to turn our results into rigorous upper bounds for the entropy.
More ambitiously, one can hope that the study of this matching problem 
will help to turn the cavity method into a rigorous tool.

\ack We thank Florent Krzakala, Thierry Mora, Martin Weigt, Guilhem Semerjian
and Federico Ricci-Tersenghi for very nice discussions and comments. This 
work has been supported by the EC trough the network MTR 2002-00319 STIPCO
and the FP6 IST consortium EVERGROW.

\end{document}